# Microwave Oscillations of a Nanomagnet Driven by a Spin-Polarised Current

S. I. Kiselev*†, J. C. Sankey*†, I. N. Krivorotov*, N. C. Emley*, R. J. Schoelkopf§, R. A. Buhrman* & D. C. Ralph*

*Cornell University, Ithaca, NY, 14853 USA

§ Department of Applied Physics and Physics, Yale University, New Haven, CT 06511 USA

† These authors contributed equally to this work

**The recent discovery that a spin-polarized electrical current can apply a large torque to a ferromagnet, through direct transfer of spin angular momentum, offers the intriguing possibility of manipulating magnetic-device elements without applying cumbersome magnetic fields.[1-16] However, a central question remains unresolved: What type of magnetic motions can be generated by this torque? Theory predicts that spin transfer may be able to drive a nanomagnet into types of oscillatory magnetic modes not attainable with magnetic fields alone,[1-3] but existing measurement techniques have provided only indirect evidence for dynamical states.[4,6-8,12,14-16] The nature of the possible motions has not been determined. Here we demonstrate a technique that allows direct electrical measurements of microwave-frequency dynamics in individual nanomagnets, propelled by a DC spin-polarised current. We show that in fact spin transfer can produce several different types of magnetic excitations. Although there is no mechanical motion, a simple magnetic-multilayer structure acts like a nanoscale motor; it converts energy from a DC electrical current into high-frequency magnetic rotations that might be applied in new devices including microwave sources and resonators.**

We examine samples made by sputtering a multilayer of 80 nm Cu / 40 nm Co / 10 nm Cu / 3 nm Co / 2 nm Cu / 30 nm Pt onto an oxidized silicon wafer and then milling through part of the multilayer (Fig. 1a) to form a pillar with an elliptical cross section of lithographic dimensions 130 nm × 70 nm.[17] Top contact is made with a Cu electrode. Transmission or reflection of electrons from the thicker "fixed" Co layer produces a spin-polarised current that can apply a torque to the thinner "free" Co layer. Subsequent oscillations of the free-layer magnetization relative to the fixed layer change the device resistance[18] so, under conditions of DC current bias, magnetic dynamics produce a time-varying voltage (with typical frequencies in the microwave range). If the oscillations were exactly symmetric relative to the direction to the fixed-layer moment, voltage signals would occur only at multiples of twice the fundamental oscillation frequency, $f$. To produce signal strength at $f$, we apply static magnetic fields ($H$) in the sample plane a few degrees away from the magnetically-easy axis of the free layer. All data are taken at room temperature, and by convention positive current $I$ denotes electron flow from the free to the fixed layer.

In characterization measurements done at frequencies < 1 kHz, the samples exhibit the same spin-transfer-driven changes in resistance reported in previous experiments[7,9] (Fig. 1b). For $H$ smaller than the coercive field of the free layer ($H_c \sim 600$ Oe), an applied current produces hysteretic switching of the magnetic layers between the low-resistance parallel (P) and high-resistance antiparallel (AP) states. Sweeping $H$ can also drive switching between the P and AP states (Fig 1b, inset). For $H$ larger than 600 Oe, the current produces peaks in the differential resistance $dV/dI$ that have been assumed previously to be associated with dynamical magnetic excitations.[4,6-8] The resistance values displayed in Fig. 1b include a lead resistance of ~ 6 Ω from high-frequency (50 GHz) probes and a top-contact resistance of ~ 9 Ω.

We measure the spectra of microwave power that result from magnetic motions by using a heterodyne mixer circuit[19] (Fig. 1a). This circuit differs from the only previous experiment to probe spin-transfer-driven magnetic oscillations[8] in that the sample is not exposed to a large high-frequency magnetic field that would alter its dynamics. The filter on the output of our mixer passes 25–100 MHz, giving a frequency resolution of ~ 200 MHz. We calibrate the circuit by measuring temperature-dependent Johnson noise from test resistors. When we state values of emitted power, they will correspond to the power that can be delivered to a load matched to the sample resistance.

We first consider the microwave spectrum from sample 1 for $H = 2$ kOe. For both negative $I$ and small positive $I$ we measure only frequency-independent Johnson noise. We will subtract this background from all the spectra we display. At $I = 2.0$ mA, we begin to resolve a microwave signal at 16.0 GHz (Fig. 1c,d). A second-harmonic peak is also present (Fig. 1c, inset). As $I$ is increased, these initial signals grow until $I \sim 2.4$ mA, beyond which the dynamics change to a different regime (Fig. 1d). In Fig. 1e, we compare the $H$-dependence of the measured frequency for the initial signals to the formula for small-angle elliptical precession of a thin-film ferromagnet[20]

$$f = \frac{\gamma}{2\pi}\sqrt{(H + H_{an} + H_d)(H + H_d + 4\pi M_{eff})}. \qquad (1)$$

Here $\gamma$ is the gyromagnetic ratio, $H_{an}$ is the within-plane anisotropy, $H_d$ models the coupling from the fixed layer, and $M_{eff}$ is the saturation magnetization minus anisotropy terms.[21] Fitting gives reasonable[22] values $4\pi M_{eff} = 8.0 \pm 0.5$ kOe, and $H_d + H_{an} = 1.18 \pm 0.04$ kOe. We therefore identify the initial signals as arising from small-angle elliptical precession of the free layer, thereby confirming pioneering predictions that spin-transfer can coherently excite this uniform spin-wave mode.[2] We can make a rough estimate for the amplitude of the precession angle, $\theta_{max}$, and the misalignment $\theta_{mis}$ between the precession axis and the fixed-layer moment (induced by the applied field) based on the integrated microwave power measured about $f$ and $2f$ ($P_f$ and $P_{2f}$). Assuming for simplicity that the angular variations are small and approximately sinusoidal in time, we calculate

$$\theta_{max}^4 \approx \frac{512 P_{2f} R_0}{\Delta R_{max}^2 I^2} \tag{2}$$

$$\theta_{mis}^2 \approx \frac{32 P_f R_0}{\Delta R_{max}^2 I^2 \theta_{max}^2}, \tag{3}$$

where $R_0 = 12.8\ \Omega$ is the device resistance and $\Delta R_{max} = 0.11\ \Omega$ is the resistance change between P and AP states. For the spectrum from sample 1 in the inset to Fig. 1c, we estimate that $\theta_{mis} \sim 9°$, and the precessional signal first becomes measurable above background when $\theta_{max} \sim 10°$.

With increasing currents, the nanomagnet exhibits additional dynamical regimes. As $I$ is increased beyond 2.4 mA to 3.6 mA for sample 1, the microwave power grows by two orders of magnitude, peak frequencies shift abruptly, and the spectrum acquires a significant low-frequency background (Fig. 1c). In many samples (including sample 2 below) the background becomes so large that some spectral peaks are difficult to distinguish. Within this large-amplitude regime, peaks shift down in frequency with increasing current (Fig. 1f). The large-amplitude signals persist for $I$ up to 6.0 mA, where the microwave power plummets sharply at the same current for which there is a shoulder in $dV/dI$. The state that appears thereafter has a DC resistance 0.04 $\Omega$ lower than the AP state and 0.07 $\Omega$ above the P state. At even higher current levels (not shown), we sometimes see additional large microwave signals that are not reproducible from sample to sample. These might be associated with dynamics in the fixed layer.

The regions of *I* and *H* associated with each type of dynamical mode can be determined by analysing the microwave power and *dV/dI* (Figs. 2a and 2b, for sample 2). In all eight samples that we have examined in detail, large microwave signals occur for a similarly-shaped range of *I* and *H*. Samples 1 and 2 exhibit clear structure in *dV/dI* at the boundaries of the large-amplitude regime, but other samples sometimes lack prominent *dV/dI* features over part of this border. In Fig. 2c we construct a dynamical stability diagram showing the different modes that can be driven by a DC spin-transfer current and a constant in-plane magnetic field. Explaining the existence of all these modes and the positions of their boundaries will provide a rigorous testing ground for theories of spin-transfer-driven magnetic dynamics.

As indicated in Figs. 2c and 2d, microwave signals can sometimes be observed not only at large *H* where dynamical modes have been postulated previously,[4,6-8,12,14-16] but also in the small-*H* regime of current-driven hysteretic switching. While sweeping to increasing currents at *H* = 500 Oe, for example, microwave peaks corresponding to small-angle precession exist for *I* within ~ 0.7 mA below the current for P to AP switching. Similar features are also observed prior to switching from AP to P at negative bias. We suggest that these microwave signals are due to fluctuations of the free-layer moment away from its easy axis to angles large enough to produce measurable precession, but too small to achieve full reversal over the activation barrier for switching.[23]

In order to understand what type of motions may be associated with the different dynamical modes, we have computed solutions of the Landau-Lifshitz-Gilbert equation of motion for a single-domain magnet.[24-27] We employ the form of the spin-transfer torque derived in [1]. The calculated zero-temperature dynamical phase diagram is presented in Fig. 3a. We have not attempted to adjust parameters to fit our data, but nevertheless the existence and relative positions of the P, AP, and small-angle-precession regimes agree well. The model suggests that the large-amplitude microwave signals correspond to large-angle, approximately-in-plane precession of the free-layer moment. The simulation reproduces the abrupt jump to much lower frequency at the onset of this mode, decreasing frequency with further increases in current (Fig. 3b), and large powers in the harmonics. The maximum simulated microwave powers for this mode in the 0-18 GHz bandwidth are 18 pW/mA$^2$ for sample 1 and 75 pW/mA$^2$ for sample 2 (differing primarily because of different $\Delta R_{max}$ values), while the measured maxima are 10 pW/mA$^2$ and 90 pW/mA$^2$, respectively. Low-frequency backgrounds in the large-amplitude spectra (e.g., Fig. 2d, spectrum 5) might be caused by fluctuations from the large-angle precessional orbit to other modes nearby in energy.[27] The single-domain simulation does not explain state W in Fig. 2c, but instead for that region it predicts approximately circular out-of plane precessional modes. These would produce large microwave signals (~25-100 pW/mA$^2$), orders of magnitude larger than the residual signals observed in state W. We suspect that our single-domain approximation may become invalid in regime W due to dynamical instabilities,[28,29] so that different regions of the sample may move incoherently, giving total time-dependent resistance changes much smaller than for single-domain motion.

The microwave power generated by the precessing nanomagnet in our devices can be quite significant. For sample 1, the largest peak in the power spectrum has a maximum more than 40 times larger than room-temperature Johnson noise. Nanomagnets driven by spin-polarised currents might therefore serve as nanoscale microwave sources or oscillators, tunable by *I* and *H* over a wide frequency range.

**Acknowledgements** We thank K. W. Lehnert, I. Siddiqi, and other members of the groups of R. J. Schoelkopf, D. E. Prober, and M. H. Devoret for advice about microwave measurements. We acknowledge support from DARPA through Motorola, from the Army Research Office, and from the NSF/NSEC program through the Cornell Center for Nanoscale Systems. We also acknowledge use of the NSF-supported Cornell Nanofabrication Facility/NNUN.

**Competing Interests Statement** The authors declare that they have no competing financial interests.

Correspondence and requests for materials should be addressed to DCR. (e-mail: ralph@ccmr.cornell.edu)**Figure 1** Resistance and microwave data for sample 1. **a**, Schematic of the sample with copper layers (orange), cobalt (blue), platinum (green), and SiO$_2$ insulator (grey), together with the heterodyne mixer circuit. Different preamplifiers and mixers allow measurements over 0.5-18 GHz or 18-40 GHz. **b**, Differential resistance versus current for magnetic fields of 0 (bottom), 0.5, 1.0, 1.5, 2.0, and 2.5 kOe (top), with current sweeps in both directions. At $H$ = 0, the switching currents are $I_c^+$ = 0.88 mA and $I_c^-$ = -0.71 mA, and $\Delta R_{max}$ = 0.11 Ω between the P and AP states. Coloured dots on the 2 kOe curve correspond to spectra shown in **c**. **b**, inset, Magnetoresistance near $I$ = 0. **c**, Microwave spectra (with Johnson noise subtracted) for $H$ = 2.0 kOe, for $I$ = 2 mA (bottom), 2.6, 3.6, 5.2, and 7.6 mA (top). We plot power density divided by $I^2$ to facilitate comparisons of the underlying changes in resistance at different current values. **c**, inset, Spectrum at $H$ = 2.6 kOe and $I$ = 2.2 mA, for which both $f$ and $2f$ peaks are visible on the same scan. **d**, Microwave spectra at $H$ = 2.0 kOe, for current values from 1.7 to 3.0 mA in 0.1 mA steps, showing the growth of the small-amplitude precessional peak and then a transition in which the second harmonic signal of the large-amplitude regime appears. **e**, Magnetic-field dependence of the small-amplitude signal frequency

(top) and the frequency of the fundamental in the large-amplitude regime at $I$ = 3.6 mA (bottom). The line is a fit to Eq. (1). **f**, Microwave power density (in colour scale) versus frequency and current for $H$ = 2.0 kOe. The black line shows $dV/dI$ versus $I$ from **b**. The curves in **b**, **c**, and **d** are offset vertically.

**Figure 2** Resistance and microwave data for sample 2, which has, at $H = 0$, $I_c^+$ = 1.06 mA, $I_c^-$ = -3.22 mA, parallel-state resistance (including top-contact and lead resistances) 17.5 Ω, and $\Delta R_{max}$ = 0.20 Ω between the P and AP states. **a**, Microwave power above Johnson noise in the frequency range 0.1–18 GHz, plotted in colour scale versus $I$ and $H$. $I$ is swept from negative to positive values. These data were collected without the mixer circuit by measuring the power with a detector diode after amplification. $P_{JN}$ is the Johnson-noise power level, and the dotted white line shows the position of the AP to P transition when $I$ is swept positive to negative. **b**, Differential resistance plotted in colour scale for the same region of $I$ and $H$. A smooth current-dependent, $H$-independent background (similar to that of Fig. 1b) is subtracted to better display the different regimes of resistance. Resistance changes are measured relative to the parallel state. **c**, Room-temperature experimental dynamical stability diagram extracted from **a** and **b**. P indicates parallel orientation, AP antiparallel orientation, P/AP parallel/antiparallel bistability, S the small-amplitude precessional regime, L the large-amplitude dynamical regime, and W a state with resistance between P and AP and only small microwave signals. The coloured dots in **c** correspond to the microwave spectra at $H$ = 500 and 1100 Oe shown in **d**.

**Figure 3** Results of numerical solution of the Landau-Lifshitz-Gilbert equation for a single-domain nanomagnet at zero temperature with the parameters[1]: $4\pi M_{eff}$=10

kOe, $H_{an}$ = 500 Oe, Gilbert damping parameter $\alpha$=0.014, and effective polarization $P$=0.3, which produce $H_c$ = 500 Oe and $I_c^+$= 2.8 mA. **a**, Theoretical dynamical stability diagram. The pictures show representative orbits for the free-layer moment vector. **b**, Dependence of frequency on current in the simulation for $H$ = 2 kOe. Black denotes the fundamental frequency, and colours correspond to harmonics in the measurement range.

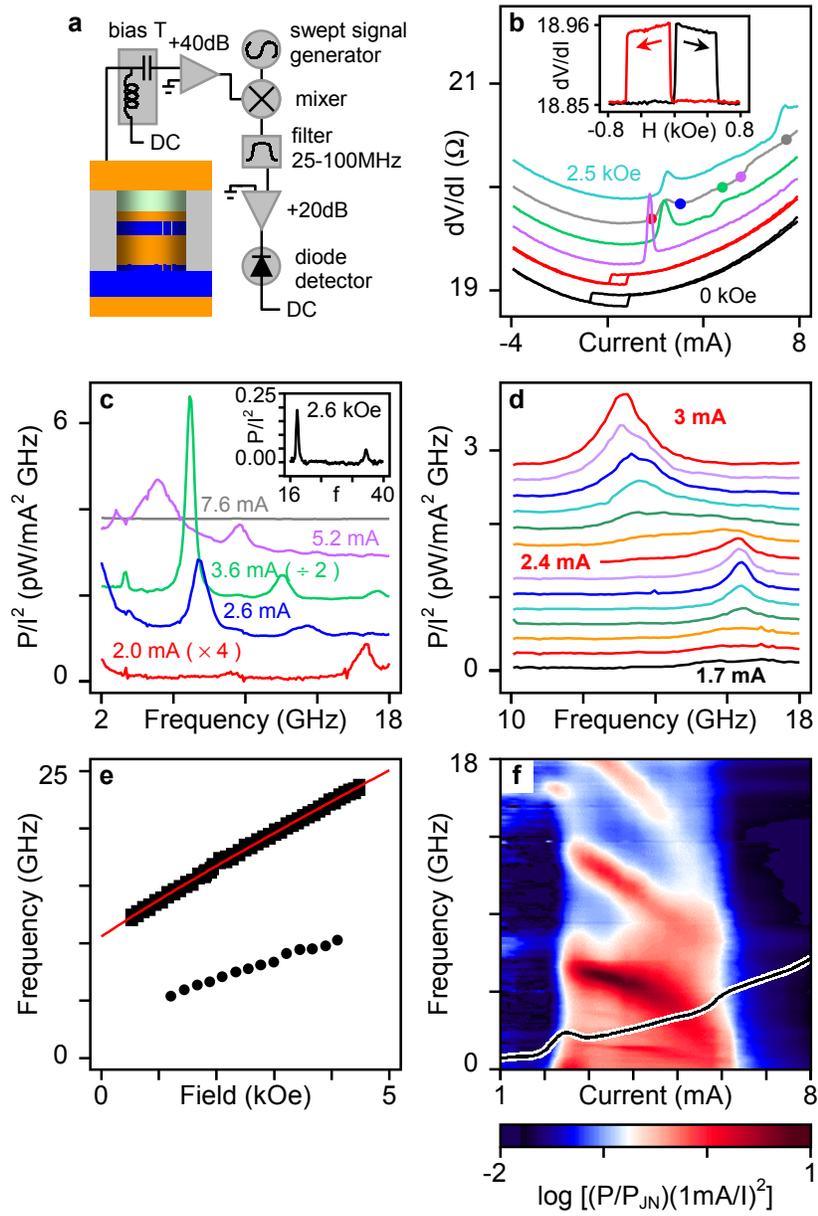

**Figure 1**

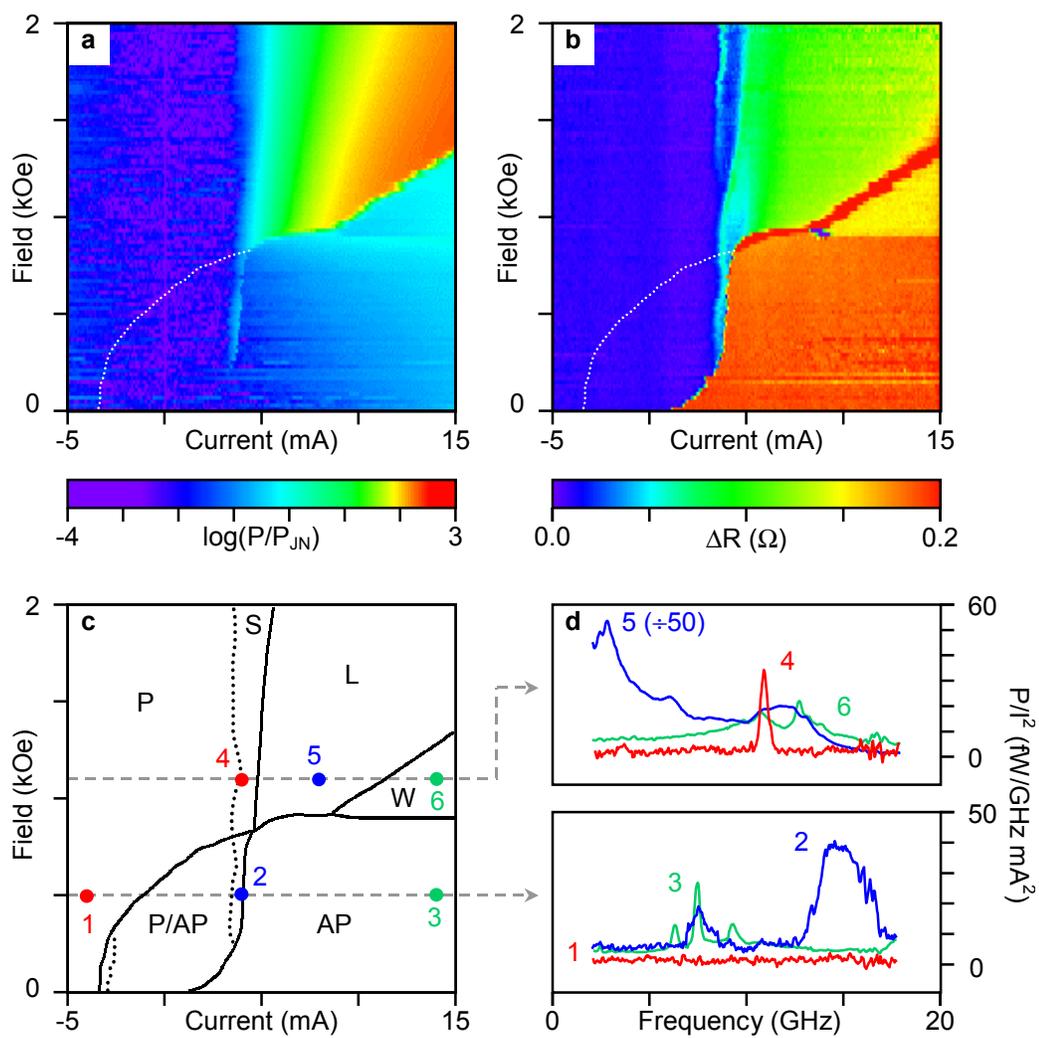

**Figure 2**

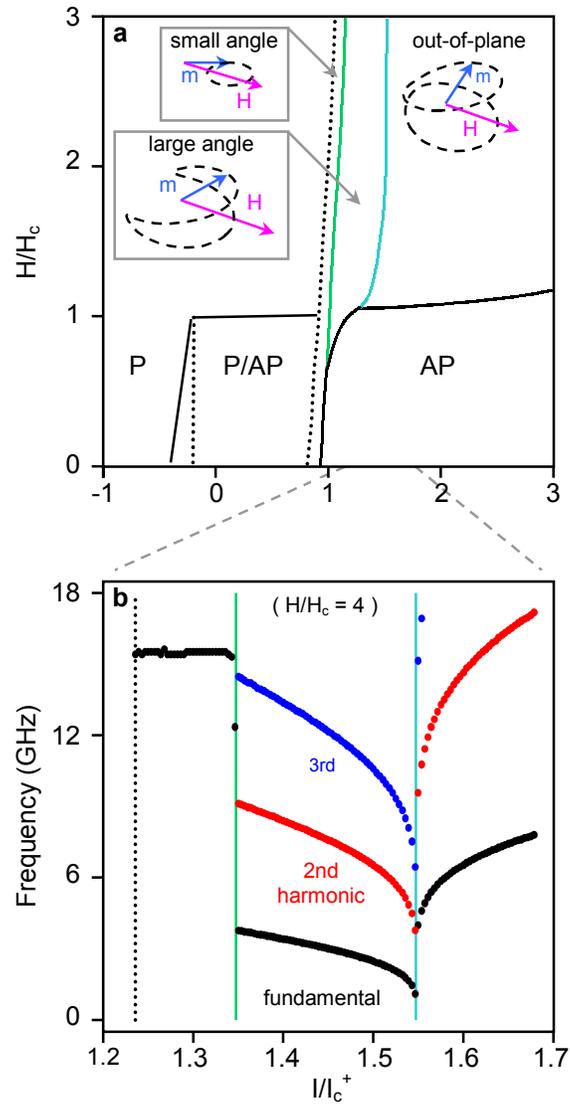

**Figure 3**